# Vortices in Bose-Einstein Condensed Na atoms

Keshav N. Shrivastava

School of Physics, University of Hyderabad,
Hyderabad 500046, India.

There are surface modes on the Bose-Einstein condensed Na atoms so that the number of vortices diverges when the stirring frequency becomes equal to that of the surface waves. We introduce the finite life time of the surface modes so that the number of vortices becomes finite. Usually the number of vortices is a linear function of the stirring frequency. We find that this linearity is destroyed by the finite life time and a peaked function emerges with several peaks, one for each surface mode. The vortices become normal, as they should be, so that there occurs a phase transition from normal to the superfluid state.

Email: keshav@mailaps.org, Fax: +91-40-3010227
Phone: +91-40-3010500 ext 4366.

______________________________________________________________

1. Introduction.

When crossed laser beams are shining on atoms, there is so much electromagnetic force that atoms become stationary. The velocity of the atoms is used to define the temperature, $(1/2)mv^2 = 3k_B T$, where a factor of three has been introduced because of the three components of the velocity. At a temperature of a few micro Kelvin, there is a Bose-Einstein condensation. It is possible to produce a stirring frequency of the order of, say, 40 Hz which can be in tune with the frequency of the "surface modes". Once the condensate is formed, the switching off of the laser beams results into the Na cloud arriving to another equilibrium in which vortices are formed[1].

In this letter, we show that the surface modes must have a complex frequency associated with a life time, otherwise the number of vortices within the available area becomes infinite.

2. Divergence.
   In the experiment of Raman et al[1], the Na atoms condense in an area of radius, R. The stirring or the rotation frequency is $\Omega$.

The vortex lines are distributed with a uniform areal density, $n_v = 2\Omega/\kappa$ where $\kappa = h/m$ with h as the Planck's constant and m the mass of one atom. The number of vortices per unit area can also be written from the vortex velocity, v and the de Broglie wave length, $\lambda$ as $n_v = 2\Omega/(v\lambda)$ where $\lambda v = h/m$ is a constant. The area of one vortex is $\kappa/(2\Omega)$. The total available area is $\pi R^2$. Therefore the number of vortices within this area is,

$$N_v = \pi R^2 / [\kappa/(2\Omega)]. \qquad (1)$$

Actually the experimental data does not agree with this simple picture of a rotational frequency $\Omega$ which can be used to determine the number of vortices in a circle of radius given by the Thomas-Fermi radius, $R = R_{TF}$. The expression (1) shows that $N_v$ is a linear function of $\Omega$, whereas the data shows, all values below the straight line and there are at least three peaks. Due to rotation, there is a centrifugal potential $-(1/2)m(\Omega r)^2$ so that the trapping frequency becomes, $(\omega_r^2 - \Omega^2)^{1/2}$ where $\omega_r$ is the mean radial frequency. The Thomas-Fermi radius is increased to a modified value,

$$R_{TF} = R_o/[1 - (\Omega/\omega_r)^2]^\alpha \qquad (2)$$

with $\alpha = 3/10$, due to rotations, so that the number of vortices becomes,

$$N_v = 2\pi\Omega\kappa^{-1}R_o^2/[1 - (\Omega/\omega_r)^2]^{2\alpha}. \qquad (3)$$

When $\Omega = \omega_r$, the denominator is zero so that the number of vortices within the given area becomes infinity. This is of course not possible. Therefore, we say that there is an unphysical divergence in the number of vortices.

3. Surface mode life time.

We make an effort to solve this problem. In eq.(1), $\Omega$ may resonate with another frequency so that at resonance, the Thomas-Fermi radius becomes infinite and hence $N_v$ in (3) diverges. There are surface mode resonances. It has been found [2-4] earlier that surface waves such as ripplons in liquid helium interact with atoms and create finite life time. Accordingly, we assume that Na atoms interact with their own surface modes. This interaction results into a self energy per Na atom which is a complex quantity. The real part gives a small shift in the energy of one Na atom and the imaginary part gives a life time. The life time of Na atoms then becomes finite. The life time of Na atoms interacting with surface modes may be taken as $\tau$. The effect of this life time is to remove the divergence of (2). The corrected Thomas-Fermi radius can therefore be written as,

$$R_{TF}^* = R_o\omega_r^{2\alpha}/[(\omega_r^2 - \Omega^2)^\alpha + (1/2\tau)^{2\alpha}] \qquad (4)$$

where we have made use of the uncertainty relation, $\Delta E.\Delta t = \eta/2$. This means that the divergence in the Thomas-Fermi radius is avoided because of the quantum mechanical uncertainty in determining the variables in the interaction between Na atoms and the surface modes. If $\tau$ is $\infty$, the correction term in the denominator of (4) becomes zero and the radius diverges. The effect of finite $\tau$ is to prevent the divergence and reduce the radius to a finite value.

We eliminate $\kappa$ from (1) to find the number of vortices as,

$$N_v = \pi R^2/[h/(2m\Omega)] \ . \qquad (5)$$

If we can measure the rotational frequency, $\Omega$ and measure the number of vortices per unit area, $N_v/\pi R^2$, then we can measure the mass m of an atom of Na. This is the equivalent of the flux quantization [5],

$$\Omega(\pi R^2) = N_v(h/2m) \qquad (6)$$

in which charge is replaced by mass. However, due to interaction with the surface ripples, the above is corrected to,

$$N_v = 2\pi m\Omega R_o^2 h^{-1}\omega_r^{4\alpha}[(\omega_r^2 - \Omega^2)^\alpha + (1/2\tau)^{2\alpha}]^{-2} \ . \qquad (7)$$

Now, when $\tau \to \infty$, the above number of vortices diverges. The effect of finite $\tau$ is to prevent the divergence of number of atoms within the available area. Now, the number of vortices as a function of rotational frequency $\Omega$ shows a resonance type shape rather than linear. When $\Omega$ is varied, there is a peak at every value of $\omega_r$. A close examination of the data of Raman et al [1] also shows that the number of vortices as a function of stirring frequency shows peaks. We can find the width at half height for one of the peaks belonging to one of the values of $\omega_r$ from the data. When stirring time is 100 ms, there is a peak centered at about 38.14 Hz. The half width at half height of this peak is 5.6 Hz, the inverse of which gives $\tau = 180$ ms. Thus, the life time associated with the first resonance at $\omega_r$ is about 180 ms, which is slightly larger than the stirring time. For very good measurements, it will be desirable to increase the stirring time. Indeed, measurements are available also with 300 ms as the stirring time. In this case also, there is a peak in the number of vortices as a function of stirring frequency and it is seen that the width at half height has remained constant and hence inverse remains at 180 ms. Therefore, our half width, the inverse of which gives the life time, is independent of the stirring time.

The observation of vortices also means that the vortices are normal and hence a phase transition from the vortex state to the superfluid state is predicted. The Abrikosov vortices usually datermine the charge of the quasiparticles but the vortices defined from (7) determine the mass of an atom. The Abrikosov vortices although made of bundles of magnetic flux, melt according to the Lindemann's criterion. The present "mass vortices" should also show melting of the lattice. Such a melting is probably present in the experimental data but has not been declared by the investigators.

4. Conclusions.

We have found that there are surface modes which interact with atoms. The life time of these interactions is finite. When this life time is taken as infinite, the number of vortices within a given area diverges which is an unphysical effect. When the life time of surface modes is taken into account, the divergence in the number of vortices disappears. The number of vortices as a function of stirring frequency ceases to be a straight line and the peaked function emerges. There is a peak for every surface mode. The vortices which emerge on stirring the Bose-Einstein condensed phase interact with their own surface modes and determine the mass of one atom so that they are different from Abrikosov vortices which determine the charge of the quasiparticles. The vortices are normal and hence a superfluid phase is predicted.

5. References.